\documentstyle[prb,aps,multicol,epsf,rotate]{revtex}

\begin{document}

\draft
\date{\today}
\title{Influence of vortex--vortex interaction on critical currents 
across low--angle grain boundaries in YBa$_2$Cu$_3$O$_{7-\delta}$ thin films}
\author{J.\ Albrecht, S.\ Leonhardt and H.\ Kronm\"uller}
\address{Max--Planck--Institut f\"ur Metallforschung, 
D--70569 Stuttgart, Germany}
\maketitle

\begin{abstract}
Low--angle grain boundaries with 
misorientation angles $\theta < 5^\circ$ in optimally doped thin films 
of YBCO are investigated by magnetooptical imaging. By using a numerical 
inversion scheme of Biot--Savart's law the critical current density across 
the grain boundary can be determined with a spatial resolution of 
about 5 $\mu$m. Detailed investigation of the spatially resolved flux density
and current density data shows that the current density across the boundary
varies with varying local flux density. Combining the corresponding flux and
current pattern it is found that there exists a universal dependency of the
grain boundary current on the local flux density. A change in the 
local flux density means a variation in the flux line--flux line distance. 
With this
knowledge a model is developped that explains the flux--current relation by
means of magnetic vortex--vortex interaction.
\end{abstract}                                                 

\begin{multicols}{2}
\narrowtext

The current limiting effect of grain boundaries in high--temperature
superconductors (HTSC's) is a topic of primary importance 
for the application of these
materials. In the last 12 years many measurements have been carried out which
show that the transport current across a grain boundary exhibits an
exponential decay with increasing misorientation angle $\phi$
\cite{Dimo88,Chis91,Gros94,Poly96,Hilg96}. 
The reason for this exponential decay is a topic of ongoing discussions and
several mechanisms to explain this behavior have been taken into 
account. At the grain boundary the local
distortion of the crystal symmetry causes an array of dislocation cores in the
superconducting film. The strain field of these dislocations creates regularly 
ordered normal conducting regions\cite{Hirt68,Sutt95,Alar95}. This leads to a
reduction of the effective superconducting interface at the grain boundary
and additionally to a reduction of the order parameter in the superconducting
regions \cite{Gure98}. This reduction can be explained by a local bending of
the electronic band structure \cite{Mann97,Hilg98}. A further point of interest
is the influence of oxygen deficiency or oxygen disorder which can lead to
the appearance of localized states at the grain boundary
\cite{Gros91,Halb92,Moec93}. Considering the pinning scenario of the flux
lines located inside the grain boundary it is necessary to remark that the
vortices are anisotropic \cite{Gure93,Gure94} which leads to an enhanced
coherence length in direction of the boundary and therefore to a reduction 
of the pinning force density.
Most of this effects which lead to the observed exponential decay, however, 
play just a secondary role if one considers grain
boundaries with low misorientation angles in thin films.
It can be experimentally found that the exponential decrease of the 
current density starts above a certain threshold
angle of about $\phi_0$~=~5$^\circ$ in case of zero 
field \cite{Hilg96,Ivan91,Hein99}.
The regime of low--angle grain boundaries (LAGB's) which means in this case 
grain boundaries with misorientation angles $\phi~<~5^\circ$ can no longer be
described by a weak link behavior because the transport current densities which
can occur across these grain boundaries can reach the values of the unperturbed
film. 

In this paper we present local measurements of the critical current
density across LAGB's in thin films performed by a magnetooptical 
technique. It can be shown
that it is not feasible to characterize these grain boundaries only by a global
transport current. The critical currents, however, are very sensitive to the
local magnetic flux density conditions inside the grain boundary. By
applying an appropriate external magnetic field it can be managed that 
the current
limiting role of the grain boundary vanishes. This effect was also found in a
similar form in YBCO bulk grain boundaries \cite{Gray98}. 

\begin{figure}[hhht]
\epsfxsize=.27\textwidth
\centerline{\rotate[r]{\epsffile{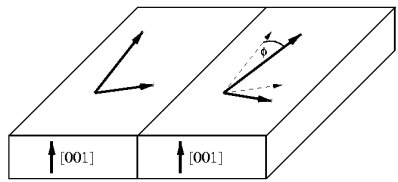}}}
\bigskip
\caption{\label{substrate} Sketch of the SrTiO$_3$--substrate geometry. The
substrates contain a symmetric [001] tilt grain boundary with a misorientation
angle $\phi~<~5^\circ$.}
\end{figure}

To investigate the role of LAGB's on the critical current density in
superconducting films the following sample geometry is used. YBCO thin films
are grown epitaxially on SrTiO$_3$ (STO) bicrystalline substrates by 
pulsed laser deposition. The STO--substrates contain a symmetric [001] tilt
grain boundary with misorientation angles $\phi~<~5^\circ$. A sketch of the
substrate geometry is shown in Fig. \ref{substrate}.   
The geometry of this grain boundary
is adapted from the superconducting film during the growth process. We want to
focus in the following on two optimally doped YBCO films with 
the dimensions of
1~mm~$\times$~1~mm~$\times$~300~nm patterned by chemical etching, which 
contain a 2$^\circ$ and a 3$^\circ$ grain boundary, respectively.

The measurements that are presented in this paper are performed by applying a
magnetooptical technique. As a field sensing 
layer a ferrimagnetic Lutetium doped Iron 
garnet film is used which is grown on a Gallium--Gadolinium garnet substrate by
liquid--phase epitaxy. This field sensing layer allows the
depiction of the magnetic flux density distribution with a spatial resolution
of 3 -- 5 $\mu$m \cite{Doro92}. The garnet 
film is observed by a polarization light
microscope and the images are obtained by a charge--coupled 
device camera with a resolution of 1000 $\times$ 1000 picture elements. 
Due to the fixed magnetic anisotropy of the indicator film a flux density range
of about 2 to 150~mT can be observed with high quantitative precision.

\begin{figure}[hhht]
\hspace*{2cm}
\epsfxsize=.65\textwidth
\centerline{\rotate[r]{\epsffile{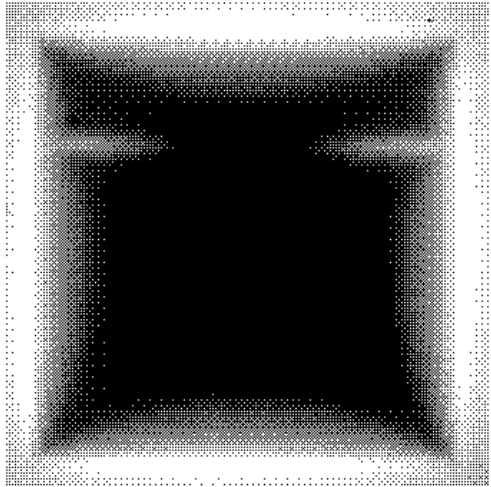}}}
\caption{\label{flux} Grayscale plot of the flux density distribution of a YBCO
film with a 3$^\circ$ grain boundary. Bright parts refer to high magnetic flux
densities. The image was obtained after zero--field cooling to 5 K with an
afterwards applied field of B$_{ex}~=~48$~mT. The sample size is 
1~$\times$~1 mm$^2$.} 
\end{figure}

In a first measurement a sample with a 3$^\circ$ grain boundary is examined by
use of the magnetooptical technique. Fig. \ref{flux} shows a grayscale image of 
the sample after
zero--field cooling (ZFC) to 5~K with an afterwards applied 
field of $B_{ex}$~=~48~mT.
Bright parts refer to high magnetic flux densities, black indicates flux--free
regions. The image shows $B_z$, the flux density component 
perpendicular to the film.  
The gray square represents the region of the superconducting
film. Magnetic flux has begun to penetrate the sample in a well--known
cushion--like form along the sample's borders \cite{Schu94}. 
The influence of the grain boundary can be seen in the two
horizontal bright lines in the upper half of the sample. These lines indicate a
large penetration of the external flux along the LAGB \cite{Albr00}. This 
is what can be expected and can be easily
understood by a reduced critical current density across the grain boundary,
which leads to an enlarged flux penetration in this region. Starting out from
the greyscale image in Fig. \ref{flux}, it can be pointed out that 
the penetration depth at the grain boundary
is about twice as large as in the unperturbed film. That means in a first order
approximation a reduced critical current density by a factor of 2 across the
LAGB.

\begin{figure}[hhht]
\begin{minipage}[t]{.24\textwidth}
\hspace*{1cm}
\epsfxsize=1.3\textwidth
\centerline{\rotate[r]{\epsffile{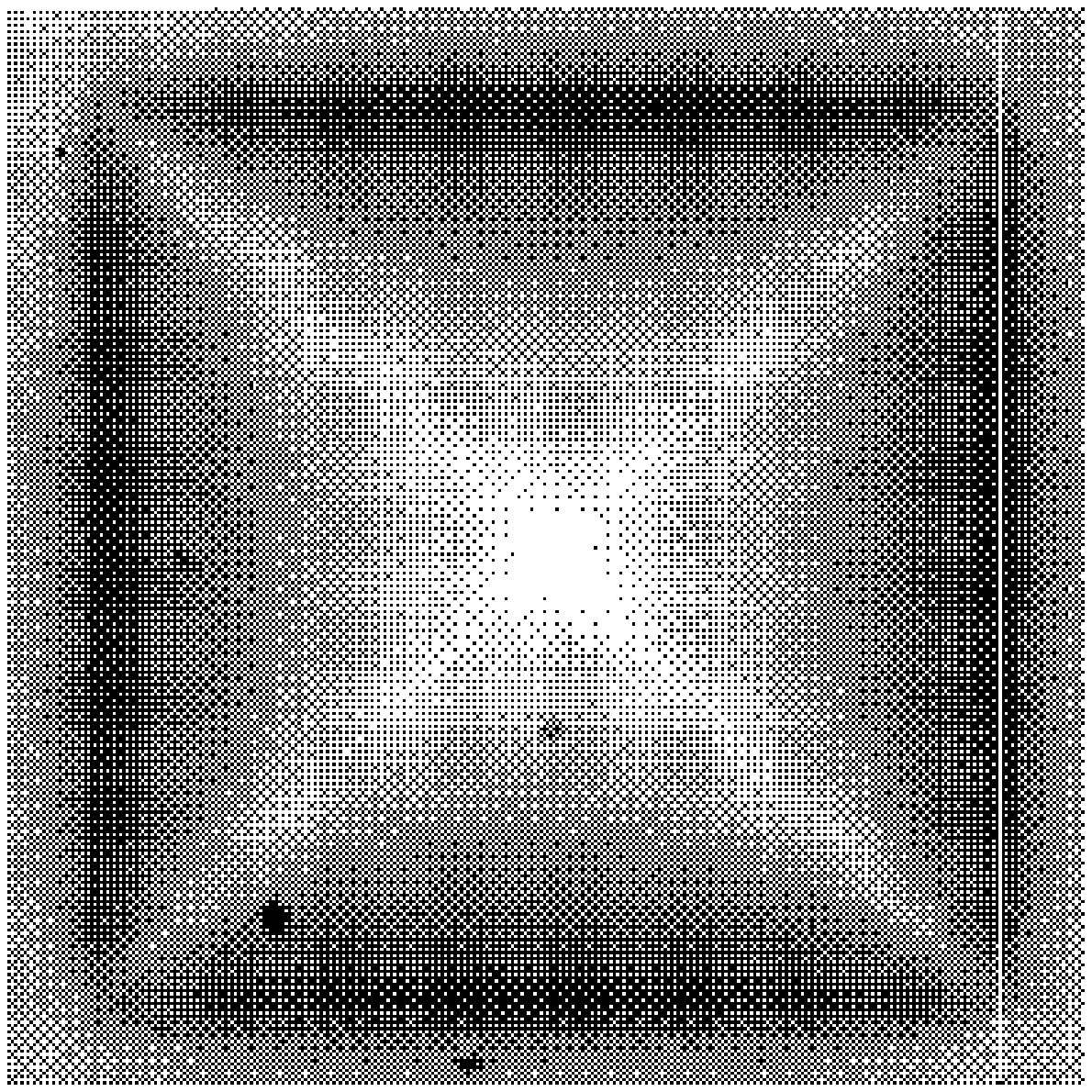}}}
\hspace*{1cm}
\epsfxsize=1.3\textwidth
\centerline{\rotate[r]{\epsffile{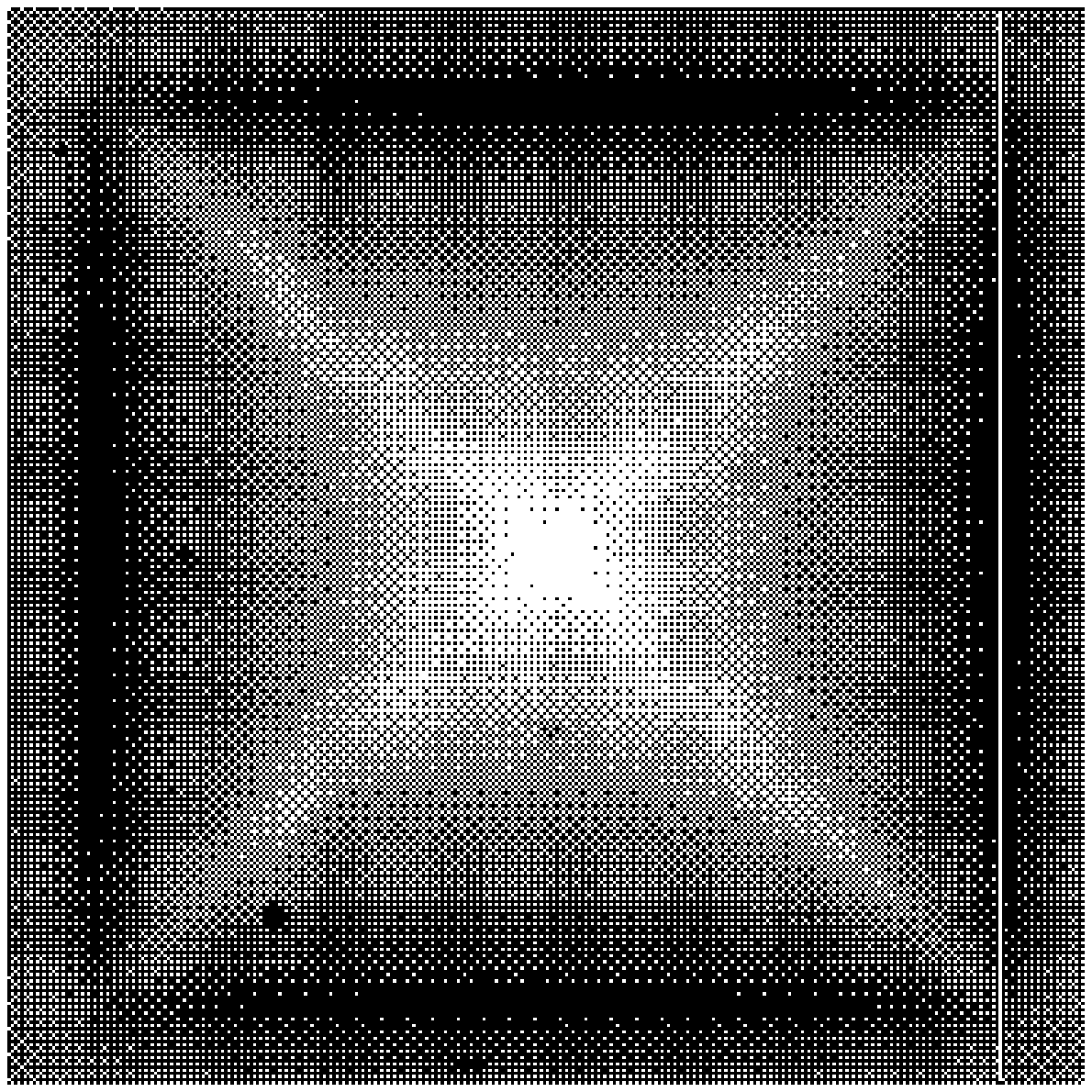}}}
\end{minipage}
\begin{minipage}[t]{.24\textwidth}
\hspace*{1cm}
\epsfxsize=1.3\textwidth
\centerline{\rotate[r]{\epsffile{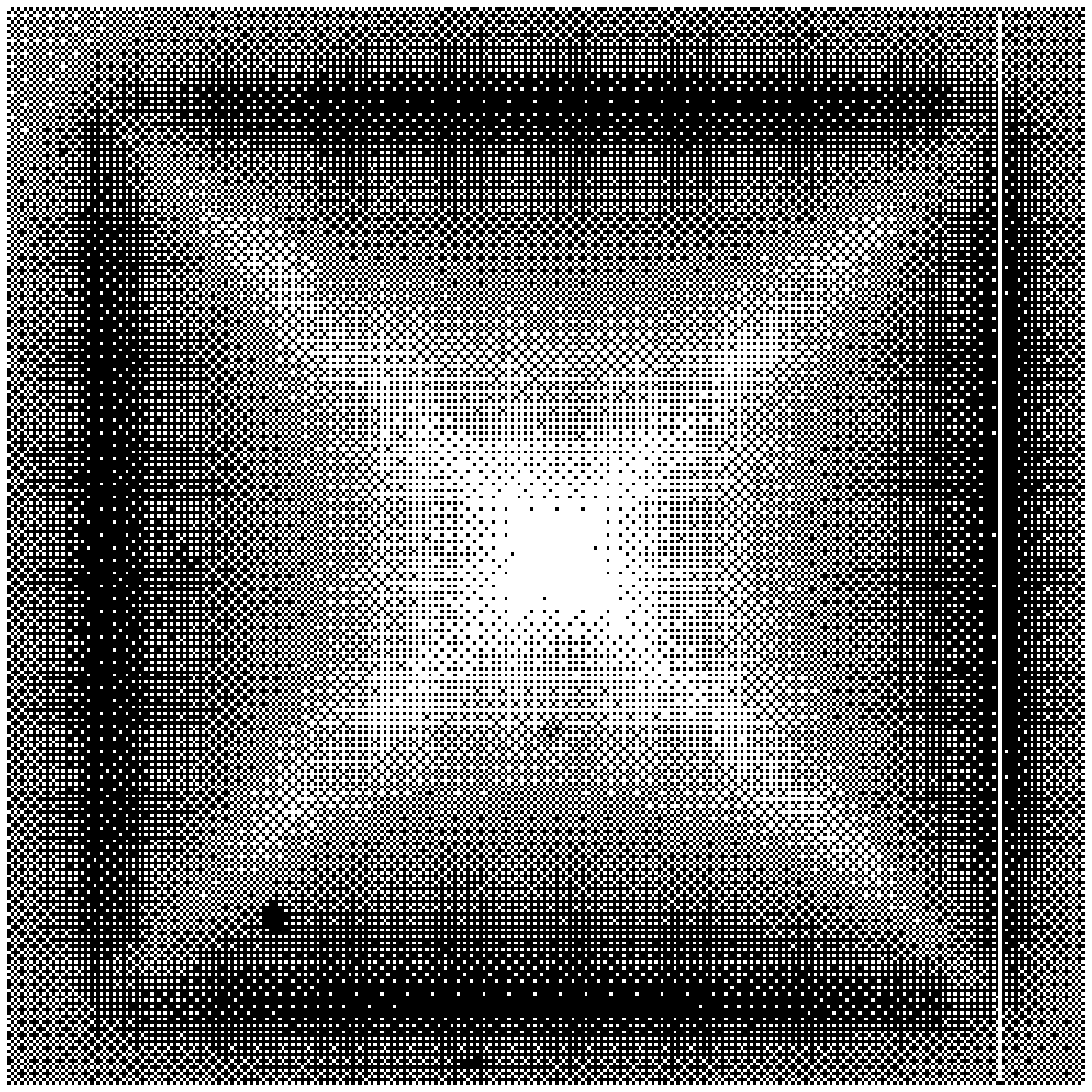}}}
\hspace*{1cm}
\epsfxsize=1.3\textwidth
\centerline{\rotate[r]{\epsffile{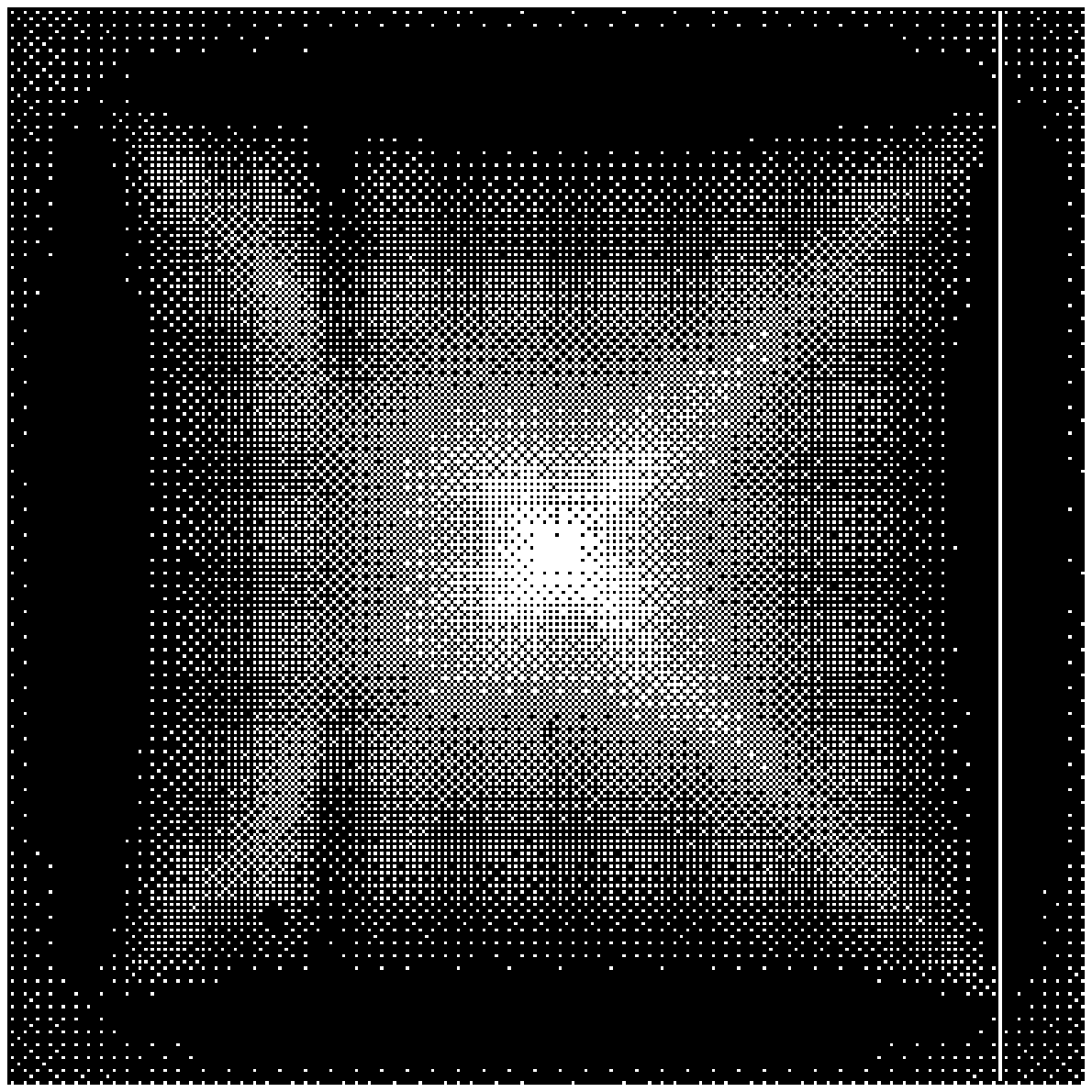}}}
\end{minipage}
\caption{\label{decrease}Magnetooptical images of the flux density distribution
of the sample in decreasing magnetic field. After applying an external magnetic
flux density of $B~\approx~500~$mT the flux density was gradually reduced. 
Shown are snapshots at 112~mT (top left), 
96~mT (top right), 80~mT (bottom left) and
48~mT (bottom right). The reappearing influence of the grain
boundary for decreasing external flux density is clearly visible as the black
lines in the upper half of the square.} 
\end{figure}

In a next step the sample is driven into the fully penetrated state by 
applying an
external magnetic flux density of $B_{ex}~\approx$~500~mT. Afterwards the
applied external field is reduced gradually. Fig. \ref{decrease} shows a series
of snapshots at B$_{ex}$~=~112~mT, 96~mT, 80~mT and 48~mT.
The behavior of the grain boundary is no longer as trivial to understand 
as for the ZFC case shown in Fig.~\ref{flux}. 
For $B_{ex}$~=112~mT, which is shown in the top left image of 
Fig.~\ref{decrease} the square--shaped film 
shows a perfect four--fold symmetry of the
flux density distribution. The white discontinuity lines \cite{Schu94} 
which indicate the
trapped flux inside the sample are exactly crossed, no perturbation by the LAGB
can be detected. This means that there exists no current limiting effect across
the grain boundary; grain boundary and unperturbed film exhibit the same
critical current. With decreasing external magnetic field, the grain boundary
reappears continuously, the influence of the grain boundary is clearly visible
again at $B_{ex}$~=~48~mT. The black lines along the LAGB indicate an
expulsion of the trapped magnetic flux due to the collapsing critical currents
across the grain boundary. These images prove that the critical current
density across a LAGB shows a strong dependence of the magnetic flux density. A
similar behavior was found for twin boundaries in YBCO single crystals
\cite{Wijn97}.
The current density increases with increasing flux density up to the value in
the unperturbed film, that means that the current limiting effect of 
an LAGB can be compensated by applying an appropriate magnetic field. 

To obtain further information about the current limiting role of LAGB's in thin
films it is necessary to determine the critical currents across the grain
boundary quantitatively. This is possible by a detailed examination of the
magnetooptical data. From the measured perpendicular 
component of the magnetic flux density $B_z$ the corresponding current density
distribution can be calculated by a numerical inversion of Biot--Savart's law.
The relation 
\begin{eqnarray}
 && B_{z}(x,y) = \nonumber \\
 && \mu_0H_{ex}+\mu_0\int_V\frac{j_x({\bf r'})(y-y')-j_y({\bf r'})(x-x')}
{4\pi|{\bf r-r'}|^3}d^3r'
\end{eqnarray} 
which is valid for a two--dimensional current density distribution 
${\bf j}=(j_x,j_y,0)$ can be inverted unambigiously by using Fourier
transformation and convolution theorem \cite{Joos98}. The lateral resolution of
the calculated current density distribution is about 5--7 $\mu m$ and is
therefore slightly reduced compared to that of the magnetic field 
data. This reduction appears because of noise effects in the measurement
\cite{Joos98}.

\begin{figure}[hhht]
\begin{minipage}[h]{.24\textwidth}
\epsfxsize=1.1\textwidth
\centerline{\rotate[r]{\epsffile{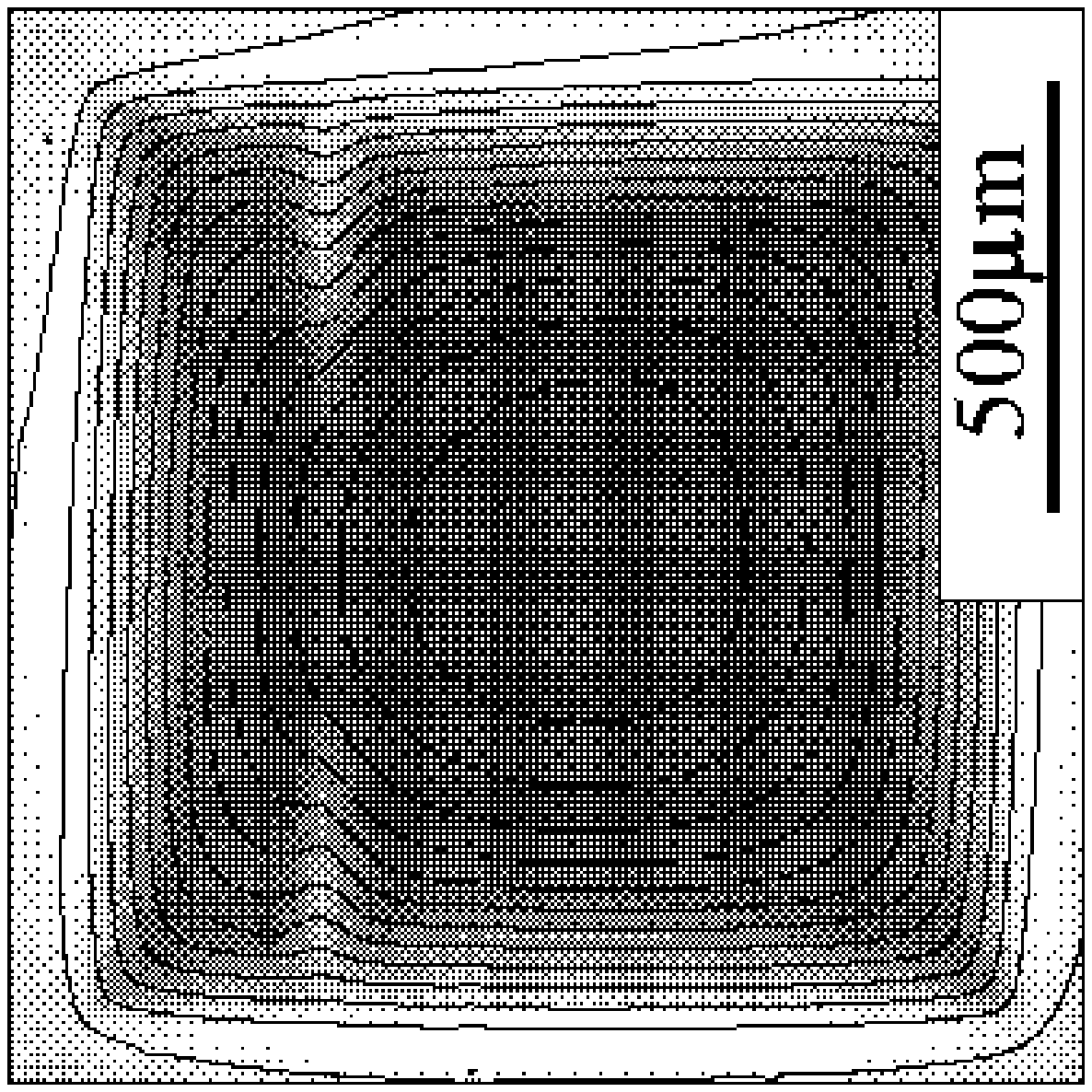}}}
\end{minipage}
\begin{minipage}[h]{.24\textwidth}
\hspace*{1cm}
\epsfxsize=1.15\textwidth
\centerline{\rotate[r]{\epsffile{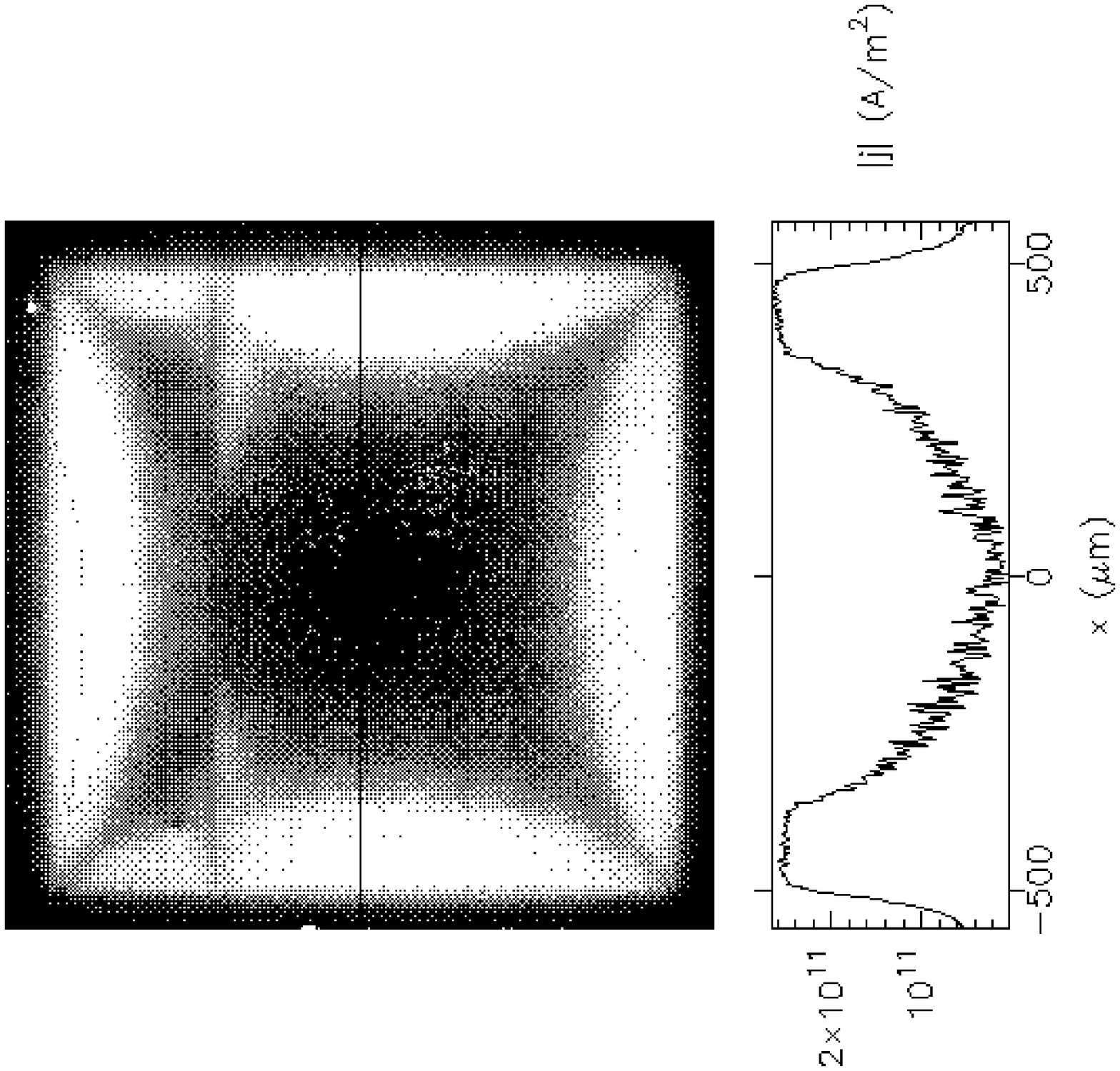}}}
\end{minipage}
\caption{\label{currents} The current density distribution calculated from the
data in Fig. \ref{flux}. The left image shows an overlay of the flux density
distribution and the calculated current stream lines, the right image shows the
absolute value of the current density, white parts refer to a high current
density.} 
\end{figure}

Two different representations of the calculated currents are 
shown in Fig. \ref{currents}.
The left image in Fig. \ref{currents} shows an overlay of the flux density
distribution known from Fig. \ref{flux} and the from these data calculated
corresponding current stream lines as solid black lines. The lines which appear
outside the sample's region are generated by numerical artefacts in the 
calculation. The influence of these virtual currents on the current pattern in
the superconducting film is very small and thus can be neglected.
An important feature of the current density distribution is the
strong bending of the stream lines in the region of the grain boundary that can 
easily be identified by the two bright lines in the upper part of the image.
In the right image the absolute value of the current density is plotted as a
grayscale. The white color indicates a current density of about
$2.5~\times~10^{11}$~A/m$^2$. This representation also shows clearly the
perturbing influence of the grain boundary in the upper half of the sample. 
The small white spots in the 
center of the sample are artefacts of the numerical calculus. A profile of the
absolute values of $j$ which is taken along the horizontal black line is
plotted below. It shows clearly the critical current in the 
flux--penetrated regions and the screening currents \cite{Bran93} in the center
of the sample. 

\begin{figure}[hhht]
\epsfxsize=.4\textwidth
\centerline{\rotate[r]{\epsffile{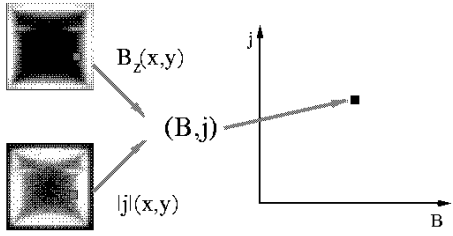}}}
\caption{\label{bj} Sketch of the procedure to obtain the local relation
between $B$ and $j$.} 
\end{figure}

The measured flux density distribution 
and the corresponding calculated current density distribution are now used 
to investigate the local relation between field and current. This investigation 
should clarify the remarkable behavior of the LAGB in the decreasing field
shown in Fig. \ref{decrease}. To obtain the local relation between flux
and current density we take the spatially resolved data from Fig. \ref{flux}
and Fig. \ref{currents}, respectively, and note down the values of flux density
and current density for every single picture element. That means for everyone
of the about 1000 $\times$ 1000 picture elements we get a couple $(B,j)$ that
can be plotted in a $B$--$j$ diagram as shown in Fig. \ref{bj}.

\begin{figure}[hhht]
\epsfxsize=.42\textwidth
\centerline{\rotate[r]{\epsffile{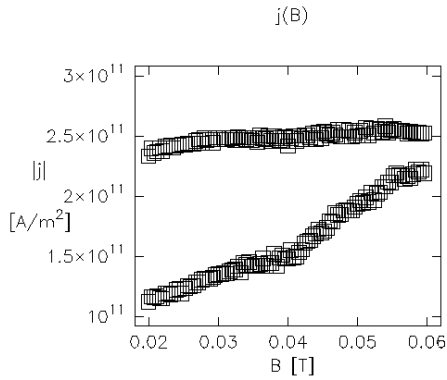}}}
\caption{\label{major} Relation between the local flux density and the 
current density. The upper curve refers to the unperturbed film, the lower
curve to the LAGB.} 
\end{figure}

This technique was now applied on two different regions of the superconductor.
First, of course, on the area of the grain boundary and second, for comparison,
on an area of the unperturbed film \cite{Albr00}. As a result we obtain the 
two curves plotted in Fig. \ref{major}.
The large difference between the two curves is obvious. The upper curve depicts
the field dependence of the critical current density in 
the unperturbed area. It shows a constant value 
of $j_c~=~2.5~\times~10^{11}$ A/m$^2$ over the considered flux density
range from 20 to 60 mT. A totally different 
behavior occurs for the $B$--$j$
relation in case of the currents across the 
grain boundary. A strong increase with
increasing flux density can be detected. The experimental data is shown in
the lower curve in Fig. \ref{major}. The curve has a nearly linear slope with 
a slight bending at $B$~=~40~mT. This bending is due to local 
variations in the microstructure of the film and will not be 
discussed any further. The increasing
current density across the grain boundary does not reach the value of the
unperturbed film in the considered flux density range, but meets 
the other curve at $B~\approx~80$ mT. This behavior goes along with the
non--perturbing influence of the LAGB in the first image of Fig.
\ref{decrease}. Note, that the flux density values of Fig. \ref{decrease} are
valid for the applied external flux density whereas in this case $B$ is the
local magnetic flux density.

\begin{figure}[hhht]
\epsfxsize=.37\textwidth
\centerline{\rotate[r]{\epsffile{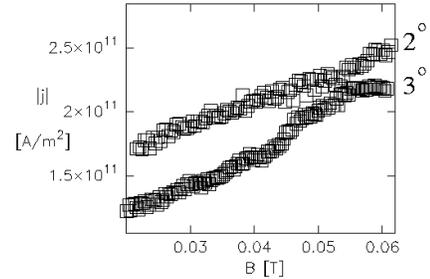}}}
\caption{\label{2and3} $B$--$j$ curves for a 2$^\circ$ (upper curve) and a 
3$^\circ$ grain boundary (lower curve). Both curves show nearly the same slope.}
\end{figure}

Fig. \ref{2and3} shows the $B$--$j$ relation for two grain boundaries with
different misorientation angles. The upper curve shows the experimental data
for a symmetric 2$^\circ$ grain boundary, the lower curve is again 
the curve from Fig. \ref{major} for comparison. 
Both of the curves show nearly the same slope, they are 
just seperated by an offset of
about 4 $\times 10^{10}$ A/m$^2$, if the small hump of the 3$^\circ$ degree
between 50 and 60~mT is neglected. This hump is probably related to a local
variation in the microstructure of the sample. Note,
that a quantitative comparison of the two measurement makes sense in this case
because both films exhibit a field independent critical current density 
of $j_c~=~2.5 \times 10^{11}$ A/m$^2$ in the unperturbed region.

The parallel shape for different angles suggests a universal field dependence
of critical currents across LAGB's which can be totally seperated from the
microstructural properties. The uniform shape of the $B$--$j$ relations is an
evidence for a additional pinning mechanism of the flux lines which 
is independent 
of the microstructural pinning of the grain boundary. Only the local magnetic
flux density and thus the flux line--flux line distance originates this effect.
As a consequence the critical current can be written as
\begin{equation}
\label{2parts}
j_c(\phi,B)=j_{c1}(\phi)+j_{c2}(B).
\end{equation}
In this equation $j_{c1}(\phi)$ represents the part of the critical current
density which is caused by the intrinsic pinning of the grain boundary. This
$j_{c1}$ shows the well known exponential decay with increasing misorientation
angle~$\phi$. $j_{c2}(B)$ has its origin only in flux line--flux line
interaction and is totally independent of the microstructure of the grain
boundary. We focus now on the contribution $j_{c2}(B)$ and try to understand 
the magnetic field dependency that we observe in our measurements.

\begin{figure}[hhht]
\epsfxsize=.3\textwidth
\centerline{\rotate[r]{\epsffile{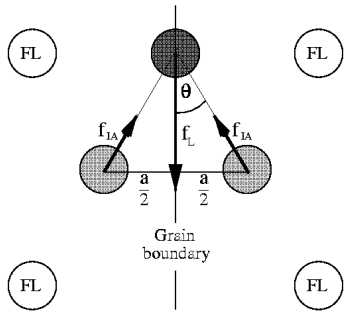}}}
\bigskip
\caption{\label{model} Sketch of the considered flux line geometry. The dark
gray circle represents a flux line located on the grain boundary. The light
gray and white circles correspond to the Abrikosov flux line lattice in
the vicinity of the grain boundary.} 
\end{figure}

To explain the shape of the B--j curve we assume a single vortex
located exactly on the LAGB and take a look at the interaction with an
Abrikosov flux line lattice in the 
vicinity of the grain boundary in presence of a Lorentz force.
A sketch of the chosen model geometry is shown in Fig. \ref{model}. 
The flux line on the
grain boundary is represented as dark gray circle at the top in 
Fig. \ref{model}, the neighboring vortices are presented light gray and white.
In the following the force per unit length shall be calculated that is 
required to drive the vortex along the grain boundary through the nearest
neighbors.  
For this calculation numerous assumptions are made which have to be discussed
first. The complex flux line--flux line interaction is reduced to 
magnetic interaction. The distance between two flux lines is several hundred
nanometers in the considered flux density range which is at least two orders of
magnitude larger than the coherence length. Therefore the flux line core
interaction can be neglected in this first order calculation. 
Also neglected is the anisotropy
of the vortex which is located exactly on the LAGB. A flux line on a 
grain boundary shows a crossover
from an isotropic Abrikosov vortex to an extremely anisotropic Josephson vortex
with increasing misorientation angle \cite{Gure93}. A certain degree of
anisotropy definitely appears in case of the LAGB's, but the fact that the
misorientation angles are very small gives rise to use isotropic vortices in
this model. In addition to this the flux lines in the unperturbed film are
assumed to be immobile and only the interaction with the nearest neighbors is
concerned. The model neglects any bending effects of the flux lines, e. g. only
the two--dimensional projection of the vortices is taken into account.

With all these restrictions the pinning contribution of this model can be
calculated. The magnetic interaction force (per unit length) between two 
vortices is given by deGennes \cite{deGe66}
$$ F_{IA}=\frac{\Phi_0^2}{2\pi\lambda^3\mu_0}K_1\left(\frac{a}{\lambda}\right).
$$
Here $\lambda$ is the London penetration depth, which is 
$\lambda\approx~150$ nm at $T~=~5$ K, $\Phi_0$ is the flux 
quantum, $a$ the flux line distance and $K_1$
the modified Bessel function or MacDonald function of first order. 
The interaction with the two
nearest neighbors in Fig. \ref{model} compensates the Lorentz force $f_L$, 
that tries to move the flux line in the LAGB towards the two light
gray flux lines. The pinning force of this geometry is now given by the maximum
force that appears, if the flux line in the LAGB is forced to pass 
through the two nearest neighbors
$$ F_{pin}=\begin{array}{c} \\ \mathrm{max}\\ \scriptstyle\mathrm{a,\theta}\\
\end{array} \left[2F_{IA}(\frac{a}{\sin\theta})\cos\theta\right], $$
$\theta$ is defined in Fig. \ref{model}.
To obtain the contribution to the critical current density across the LAGB
it is necessary to calculate the current density from the force per unit
length. Due to the fact that the magnetic interaction force is present over the
whole length of the vortex, one obtains easily $j_{c2}(a)=F_{pin}(a)/\Phi_0$,
which is the well known definition of a Lorentz force.
For better comparison to the experimental data, the dependency on the
flux line--flux line distance $j(a)$ is transformed into a flux density 
dependence $j(B)$ using $B=2\Phi_0/\sqrt{3}a^2$ for a triangular Abrikosov
lattice. 

\begin{figure}[hhht]
\epsfxsize=.42\textwidth
\centerline{\rotate[r]{\epsffile{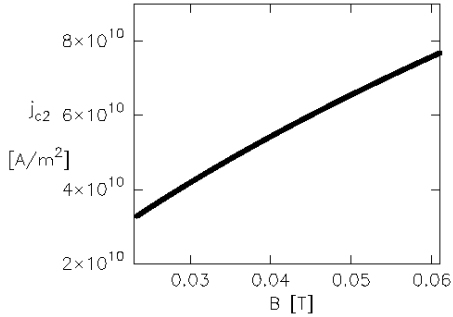}}}
\caption{\label{jc2} Contribution of the magnetic vortex--vortex interaction to
the critical current density across the grain boundary.} 
\end{figure}

The resulting relation for the interesting flux density range is plotted in
Fig. \ref{jc2}.
The plot shows a similar increase of the critical current 
density with increasing flux density as found in the experimental data and the
calculation yields current densities of the right order of magnitude.

\begin{figure}[hhht]
\epsfxsize=.42\textwidth
\centerline{\rotate[r]{\epsffile{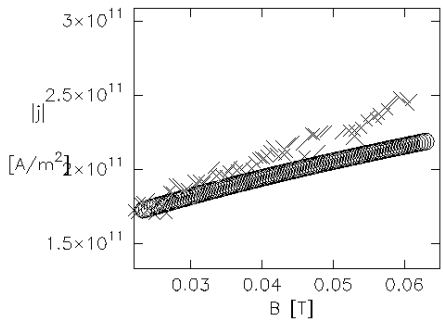}}}
\caption{\label{result} Comparison of the calculated j--B relation (circles) 
with the experimental data (crosses) of the 2$^\circ$ grain boundary.} 
\end{figure}

For an optimal comparison to the measurement the field independent part
$j_{c1}$ of the critcal current density of the grain boundary has to be 
estimated. This can be performed by comparing 
the data for very low magnetic flux
densities where the contribution $j_{c2}$ is small. Using the data below a
local flux density $B$~=~30~mT a value of $j_{c1}~=~1.4~\times~10^{11}$~A/m$^2$
fits the data best in case of the 2$^\circ$ grain boundary. Plotting now 
$j_c~=~j_{c1}~+~j_{c2}$ versus the experimental data one obtains 
Fig.~\ref{result}. 
Both the experimental and the calculated curve show a similar shape. The
measured data show a stronger increase than the model predicts but the
slope of both curves is in the same range. The largest deviation from the model
prediction is found for higher magnetic fields. A possible improvement of the
very simple model might be the consideration of more than just
nearest--neighbor interaction of the flux lines especially for the higher field
range \cite{Pear64}.

To summarize our results, the critical current density across low--angle grain
boundaries in thin films of YBCO is investigated by a high--resolution method.
The analysis of the local field dependence of the critical current shows a
uniform behavior for different misorientation angles. This uniformity can be
explained by seperating two parts that contribute to the critical 
current density. One
part is correlated to the microstructural properties of the grain boundary and
shows the typical drop for increasing misorientation angles. The second part is
independent of the microstructure and can be described by vortex--vortex
interaction in and in the vicinity of the grain boundary. A model which takes
the deGennes magnetic interaction into account is able to reproduce the
measured current densities.

The authors are grateful to Ch. Jooss, R. Warthmann and M. V. Indenbom 
for stimulating and 
helpful discussions and to G. Cristiani and H.--U. Habermeier for the 
preparation of the excellent samples.

{

}
\end{multicols}
\end{document}